\documentclass[pre,preprint, groupedaddress, showkeys,showpacs]{revtex4}
\usepackage{graphicx}
\begin{document}

%\preprint{}

\title{Transport of overdamped Brownian particles in a two-dimensional tube: Nonadiabatic regime}


\author{Bao-quan  Ai}
%\homepage[]{}

%\thanks{}
\affiliation{Laboratory of Quantum Information Technology, ICMP and
 SPTE, South China Normal University, Guangzhou, China.}


\date{\today}
\begin{abstract}
\indent Transport of overdamped Brownian particles in a
two-dimensional asymmetric tube is investigated in the presence of
nonadiabatic periodic driving forces. By using Brownian dynamics
simulations we can find that the phenomena in nonadiabatic regime
differ from that in adiabatic case. The direction of the current can
be reversed by tuning the driving frequency. Remarkably,  the
current as a function of the driving amplitude exhibits several
local maxima at finite driving frequency.
\end{abstract}

\pacs{ 05.40.-a, 07. 20. Pe}
\keywords{Brownian motor, ratchet, nonadiabatic, current reversal}



\maketitle
\section {Introduction}
\indent Rectification of noise leading to unidirectional motion in
ratchet systems has been an active field of research over the last
decade\cite{a1}. This comes from the desire to understand molecular
motors\cite{a2}, nanoscale friction\cite{a3}, surface
smoothing\cite{a4}, coupled Josephson junctions\cite{a5}, optical
ratchets and directed motion of laser-cooled atoms\cite{a6}, and
mass separation and trapping schemes at the microscale\cite{a7}. In
these systems possessing spatial or dynamical symmetry breaking,
Brownian motion combined with unbiased external input signals,
deterministic and random alike, can assist directed motion of
particles at submicron scales.

\indent Several models have been proposed to explain this transport
mechanism under various nonequlibrium situations. Typical examples
are rocking ratchets \cite{a8}, flashing ratchets\cite{a9},
diffusion ratchets\cite{a10}, correlation ratchets\cite{a11}. The
ratchet setup demands three key ingredients\cite{a12} which are (a)
nonlinearity: it is necessary since the system will produce a zero
men out put from zero-mean input in a linear system. (b) asymmetry
(spatial and /or temporal): it can violate the left /right symmetry
of the response. (c) fluctuating input zero mean force: it should
break thermodynamical equilibrium, which forbids appearance of a
directed transport due to the Second Law of Thermodynamics.

\indent Most studies have revolved around the energy barrier. The
nature of the barrier depends on which thermodynamic potential
(internal energy or Helmholtz free energy) varies when passing from
one well to the other, and its presence plays an important role in
the dynamics of the system. Whereas energy barriers are more
frequent in problems of solid-state physics (metals and
semiconductors, coupled Josephson junction, and photon crystal).
However,  in some cases, such as soft condensed-matter and
biological systems, the entropy barriers should be considered.
Brownian particles, when moving in a confined geometry, instead of
diffusing freely in the host liquid phase, undergo a constrained
motion, where their kinetic behavior could exhibit peculiar
behavior. This feature of constrained motion is ubiquitous in ion
channels, nanopores, zeolites, and generally for processes occurring
at sub-cellular level\cite{a13}. Entropic barriers may appear when
coarsening the description of a complex system in order to simplify
its dynamics. Reguera and Rubi\cite{a14} used the mesoscopic
nonequilibrium thermodynamics theory to derive the general kinetic
equation of the motor system and analyzed in detail the case of
diffusion in a domain of irregular geometry in which the presence of
the boundaries induces an entropy barrier when approaching the
dynamics by a coarsening of the description. In their recent work
\cite{a15} they studied the current and the diffusion of a Brownian
particle moving in a symmetric channel with a biased external force.
They found that temperature dictates the strength of the entropic
potential, and thus an increase of temperature leads to a reduction
of the current. In our previous work\cite{a16}, we found that the
asymmetry of the tube can induce a net current in the absence of any
net macroscopic forces or in the presence of the unbiased forces in
the adiabatic case. The present work is extended to the study of
transport to the nonadiabatic regime. We emphasize on finding how
the finite driving frequency affects the transport.

\section{Model and Methods}
\begin{figure}[htbp]
  \begin{center}\includegraphics[width=10cm,height=6cm]{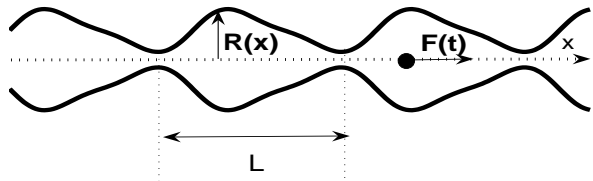}
  \caption{Schematic diagram of a tube with periodicity $L$. The
  shape is described by the radius of the tube $R(x)=a[\sin(\frac{2\pi x}{L})+\frac{\Delta}{4}\sin(\frac{4\pi
    x}{L})]+b$. $\Delta$ is the asymmetric parameter of the tube shape. $F(t)=A_{0}\sin(\omega t)$ is an external driving force.}\label{1}
\end{center}
\end{figure}
\indent  In this paper, we study a ratchet-like periodic tube driven
by a thermal noise and a periodic driving force. Since most of the
molecular transport occurs in the overdamped regime, we can safely
neglect initial effects. So, its overdamped dynamics is described by
the following Langevin equations written in a dimensionless form
\cite{a15,a16},
\begin{equation}\label{}
    \eta\frac{dx}{dt}=A_{0}\sin(\omega t)+\sqrt{\eta
    k_{B}T}\xi_{x}(t),
\end{equation}
\begin{equation}\label{}
    \eta\frac{dy}{dt}=\sqrt{\eta
    k_{B}T}\xi_{y}(t),
\end{equation}
where $x$, $y$, are the three-dimensional (2D) coordinates, $\eta$
is the friction coefficient of the particle, $k_{B}$ is the
Boltzmann constant, $T$ is the absolute temperature and
$\xi_{x,y}(t)$ is the Gaussian white noise with zero mean and
correlation function:
$<\xi_{i}(t)\xi_{j}(t^{'})>=2\delta_{i,j}\delta(t-t^{'})$ for
$i,j=x, y$. $<...>$ denotes an ensemble average over the
distribution of noise. $\delta(t)$ is the Dirac delta function.
$A_{0}$ and $\omega$ are the amplitude and frequency of the external
driving force, respectively. Imposing reflecting boundary conditions
in the transverse direction ensures the confinement of the dynamics
within the tube, while periodic boundary conditions are enforced
along the longitudinal direction for the reasons noted above. The
shape of the tube is described by its radius
\begin{equation}\label{}
    R(x)=a[\sin(\frac{2\pi x}{L})+\frac{\Delta}{4}\sin(\frac{4\pi
    x}{L})]+b,
\end{equation}
where $a$ is the parameter that controls the slope of the tube,
$\Delta$ the asymmetry parameter of the tube shape.  $b$ is the
parameter that determine the radius at the bottleneck.

\indent If $F(t)$ changes very slowly with respect to $t$, namely,
its period is longer than any other time scale of the system,
 there exists a quasisteady state. In adiabatic limit and  $|R^{'}(x)|<<1$, by following the  method in  \cite{a14,a15}, we can obtain the
current
\begin{equation}\label{}
    j(F(t))=\frac{k_{B}T[1-\exp(-\frac{F(t)L}{k_{B}T})]}{\int_{0}^{L}\int_{0}^{L}dxdy \frac{R(x)}{R(x+y)}\exp[\frac{-F(t)y}{k_{B}T}][1+R^{'}(x+y)^{2}]^{\alpha}},
\end{equation}
where $\alpha=1/3$ and the prime stands for the derivative with
respect to the space variable $x$. So the average current
\cite{a15,a16} is
\begin{equation}\label{9}
    J=\frac{1}{\tau}\int_{0}^{\tau}j(F(t))dt=\frac{1}{2}[j(A_{0})+j(-A_{0})],
\end{equation}
and the average velocity $\nu=JL$.

\indent Our model can be analytically studied in the adiabatic
limit\cite{a15}. However, in present work we are interested in the
intermediate frequency and strong amplitude of driving force. In this
case, no general valid analytical expressions are possible.
Therefore, we use Brownian dynamic simulations performed within the
stochastic Euler-algorithm by integration of the dimensionless
Langevin equations(1-2). For the numerical simulations the single
integration steps read \cite{a17}:
\begin{equation}\label{}
    x(t+\Delta t)=x(t)+A_{0}\sin(\omega t)\Delta t+\sqrt{2k_{B}T \Delta
    t}R_{1},
\end{equation}
\begin{equation}\label{}
y(t+\Delta t)=y(t)+\sqrt{2k_{B}T \Delta t}R_{2},
\end{equation}
where $R_{1}$, $R_{2}$ are two Gaussian distributed random numbers
with unit variance. $\Delta t$ is the integration step time. If the
new desired position is not allowed in the sense that it is lying
outside the channel then the boundary conditions have to be
considered, i.e the simulation step is discarded. For the numerical
simulations, we have considered more than $1\times 10^{4}$
realizations to improve accuracy and minimize statistical errors. In
order to provide the requested accuracy of the system dynamics time
step was chosen to be smaller than $10^{-4}$. The average particle
velocity along the $x$-direction,
\begin{equation}\label{}
\nu=\langle\dot{x}\rangle=\lim_{t\rightarrow\infty}\frac{\langle
x(t)
    \rangle}{t}.
\end{equation}

\section {Numerical results and discussion}
Our emphasis is on finding the asymptotic mean velocity which is
defined as the average of the velocity over the time and thermal
fluctuations. In the nonadiabatic regime, we cannot obtain the
similar expression for the current to that in adiabatic limit [Eq.
(5)], therefore, we carried out extensive numerical simulations. For
simplicity we set $\eta=1$ and $k_{B}=1$ throughout the work.

\begin{figure}[htbp]
   \begin{center}\includegraphics[width=10cm,height=8cm]{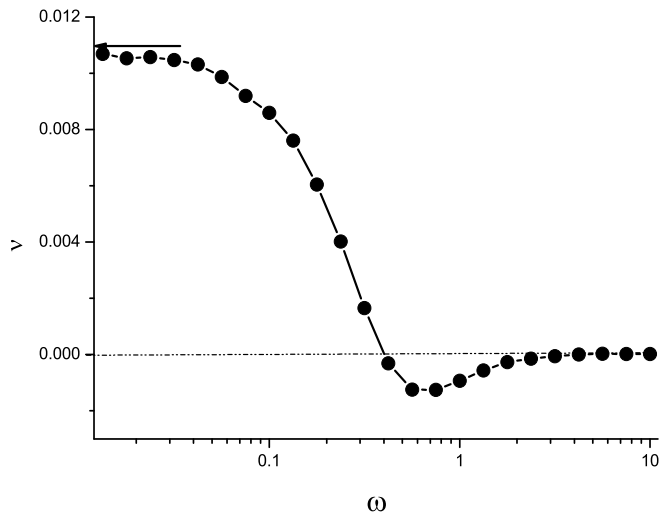}
  \caption{The mean velocity $\nu$ as a function of driving frequency $\omega$ at $a=1/2\pi$, $b=1.2/2\pi$, $A_{0}=0.5$, $\Delta=1.0$, $T=0.3$, and $L=2\pi$. The arrow marks the mean velocity calculated in the
  adiabatic limit.}\label{1}
\end{center}
\end{figure}

\indent Figure 2 shows the mean velocity $\nu$ as a function of the
driving frequency $\omega$. The only resource driving particle
current across the tube  is the nonequlibrium, external driving
force $F(t)$, which generates positive and negative driving force in
first and second half of the driving period. In the adiabatic limit
$\omega\rightarrow 0$, the external force can be expressed by two
opposite static forces $A_{0}$ and $-A_{0}$, yielding the mean
current $J=\frac{1}{2}[j(-A_{0})+j(A_{0})]$. In this case it is
easier for the particles moving towards the slanted side than
towards the steeper side, so the current is positive. On increasing
the frequency $\omega$, due to higher frequency the Brownian
particles do not get enough time to cross the slanted entropic
barrier (the right side) which is at a larger distances from the
minima. Since the distance from a entropic minima to the basin of
attraction of next minima is less than from the steeper side (the
left side) than from the slanted side (the right side, hence in one
period the particles get enough time to climb the entropic barrier
from the steeper side than from the slanted side, resulting in a
negative current.  When the external driving force changes very fast
$\omega\rightarrow \infty $, the particle will experience a time
averaged constant force $F=\int_{0}^{\frac{2\pi}{\omega}}F(t)dt=0$,
so the current tends to zero. At the adiabatic limit the values of
$\nu$ from Eq. (5) agree well with the numerical results.
Interestingly, at some intermediate value $\omega$, the current
crosses zero and subsequently reverses its direction. There exists a
valley in velocity-frequency curve. This peculiar reversal is due to
different strength and symmetry of relaxation processes. The
symmetry of relaxation processes changes with the system parameters,
so the current may changes its direction when the parameters are
changed.
\begin{figure}[htbp]
  \begin{center}\includegraphics[width=10cm,height=8cm]{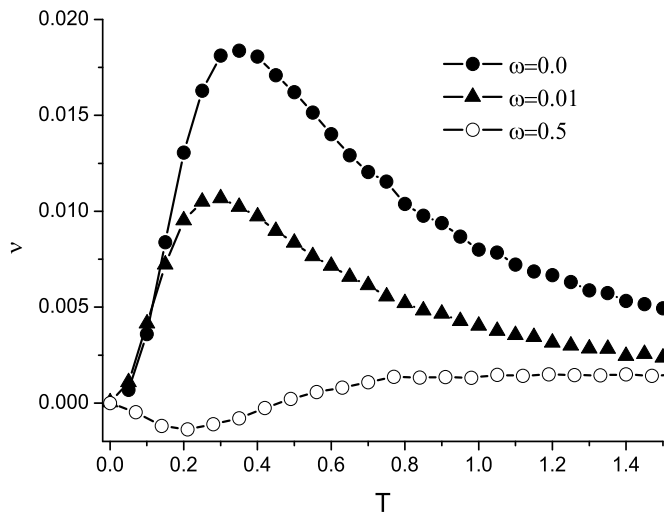}
  \caption{The mean velocity $\nu$ as a function of the temperature $T$ for different values of driving frequency $\omega$ at $a=1/2\pi$, $b=1.2/2\pi$, $A_{0}=0.5$, $\Delta=1.0$, and $L=2\pi$.}\label{1}
\end{center}
\end{figure}

\indent In Fig. 3, the mean velocity $\nu$ is plotted for several
driving frequency $\omega$ as a function of the temperature $T$. The
adiabatic limit $\omega=0$ drawn as dotted line is readily evaluated
from Eq. (5). In the determined limit $T\rightarrow 0$, the particle
cannot reach the 2D area and the effect of the asymmetry of the tube
disappears and there is no current. We must point out that in the determined limit no
currents occur even for very large amplitude driving forces which is
different from that in rocking potential ratchets \cite{c1,c2,c3}. In
the potential case, the current will occur in the determined limit
for large amplitude driving forces. When the temperature is very
high, the influence of the external driving forces becomes
negligible, so the current will also go to zero. In the adiabatic
limit, the current is always positive for $\Delta>0$. However, the
current will change its direction on increasing $T$ for finite
frequency driving force ($\omega=0.5$). In this case, at low
temperature the particles get enough time to climb the entropic
barrier from the left side and do not get enough time climb the
entropic barrier from the right side, so the current is negative. On
increasing the temperature, the particles get kicks of larger
intensity and hence they easily cross the right side, resulting in
positive current. On further increasing the temperature, the effect
of the external driving force disappear, so the current tends to
zero.

\begin{figure}[htbp]
  \begin{center}\includegraphics[width=10cm,height=8cm]{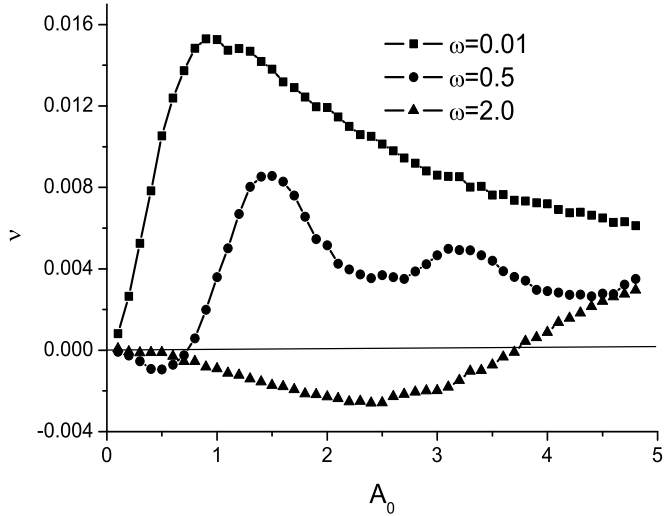}
  \caption{The mean velocity $\nu$ as a function of driving amplitude $A_{0}$ for different values of driving frequency $\omega$ at $a=1/2\pi$, $b=1.2/2\pi$, $T=0.3$, $\Delta=1.0$, and $L=2\pi$.}\label{1}
\end{center}
\end{figure}

Figure 4 shows the mean velocity $\nu$ versus the driving amplitude
$A_{0}$ for different values of $\omega$. It is found that the
current will tend to zero for very small and large amplitude driving
force. This can be understood upon noting that the driving force can
negligible for small amplitude driving forces and the effect of the
asymmetry of the tube will disappear for large amplitude driving
forces. In the adiabatic limit, the mean velocity will be always
positive for $\Delta>0$. However, this phenomena will change
drastically for nonadiabatic case. The current reversal will occur
when the amplitude of the driving force is increased. Remarkably,
one finds several extrema in velocity-amplitude characteristics
which is similar to that in periodically rocked thermal
ratchet\cite{c1,c2,c3}. This is due to the mutual interplay between
noise and finite-frequency driving forces.

\section{Concluding Remarks}
\indent In this study, we have studied the transport properties of
overdamped Brownian particles moving in an asymmetric periodic tube.
The model can be analytically studied in the adiabatic limit and
numerically in the nonadiabatic regime. We focus on finding how the
finite driving frequency affects the transport of the overdamped
particles. The phenomena in nonadiabatic regime are different from
those in adiabatic limit. From Brownian dynamics simulations, we
observes several novel and complex features arising due to the
mutual interplay between the thermal noise and the finite frequency
driving force. The directed transport is determined by two factors:
the thermal noise induced the particles escape from the well and the
external force induced relaxation processes inside the tube. It is
found that at some intermediate value of the driving frequency the
current crosses zero and subsequently reverse its direction.
Therefore, one can control the direction of the current by suitably
tailoring the frequency of the driving force. In addition one finds
several extrema in the current-amplitude characteristic.

\indent Though the model presented does not pretend to be a
realistic model for a real system, the results we have presented
have a wide application in may processes, such as catalysis, osmosis
and particle separation, and on the noise-induced transport in
periodic potential landscapes that lack reflection symmetry, such as
ratchet systems. It is very important to understand the novel
properties of these confined geometries, zeolites, biological
channels, nanoporous materials, and microfluidic devices, as well as
the transport behavior of species in these systems.

\indent  This work was supported in part by National Natural Science
Foundation of China with Grant No. 30600122 and GuangDong Provincial
Natural Science Foundation with Grant No. 06025073.

\end{document}